\documentclass[natbib]{svjour3}                     
\smartqed  
\usepackage{graphicx}
\usepackage{epsf}
\usepackage{lscape}
\newcommand{\aap}{{Astron. Astrophys.}}
\newcommand{\aj}{{Astronomical J.}}
\newcommand{\apj}{{Astrophys. J.}}
\newcommand{\apjs}{{Astrophys. J. Suppl.}}
\newcommand{\apjl}{{Astrophys. J.}}

\newcommand{\jgr}{{Journal of Geophys. Res.}}

\newcommand{\nat}{{Nature}}
\newcommand{\araa}{{Ann. Rev. Astron. Astro.}}
\newcommand{\planss}{{Plan. Space Sci.}}
\newcommand{\prb}{{Phys. Rev. B}}
\newcommand{\mnras}{{MNRAS}}
\newcommand{\mj}{$M_{\mathrm{J}}$}
\newcommand{\rj}{$R_{\mathrm{J}}$}
\newcommand{\me}{$M_{\oplus}$}

\newcommand{\hd}{HD 209458b}

\newcommand{\teff}{$T_{\rm eff}$}

\newcommand{\cp}{\citep}
\newcommand{\ct}{\citet}

\begin{document}

\title{The Interior Structure, Composition, and Evolution \\of Giant Planets}

\titlerunning{Giant Planet Interiors}        

\author{Jonathan J. Fortney\footnote{Both authors contributed equally to this work.}         \and
        Nadine Nettelmann$^{\star}$ 
}


\institute{
J.~J. Fortney \at
              Department of Astronomy and Astrophysics \\
	      University of California, Santa Cruz\\
              \email{jfortney@ucolick.org}            \\
           \and
		N. Nettelmann \at
		Institut fur Physik\\
		Universitat Rostock\\
		\email{nadine.nettelmann@uni-rostock.de}\\
}

\date{Received: date / Accepted: date}

\maketitle

\begin{abstract}
We discuss our current understanding of the interior structure and thermal evolution of giant planets.  This includes the gas giants, such as Jupiter and Saturn, that are primarily composed of hydrogen and helium, as well as the ``ice giants,'' such as Uranus and Neptune, which are primarily composed of elements heavier than H/He.  The effect of different hydrogen equations of state (including new first-principles computations) on Jupiter's core mass and heavy element distribution is detailed. This variety of the hydrogen equations of state translate into an uncertainty in Jupiter's core mass of 18$M_{\oplus}$. For Uranus and Neptune we find deep envelope metallicities up to 0.95, perhaps indicating the existence of an eroded core, as also supported by their low luminosity.  We discuss the results of simple cooling models of our solar system's planets, and show that more complex thermal evolution models may be necessary to understand their cooling history.  We review how measurements of the masses and radii of the nearly 50 transiting extrasolar giant planets are changing our understanding of giant planets.  In particular a fraction of these planets appear to be larger than can be accommodated by standard models of planetary contraction.  We review the proposed explanations for the radii of these planets.  We also discuss very young giant planets, which are being directly imaged with ground- and space-based telescopes.
\end{abstract}

\section{Introduction}
\label{intro}
In order to understand the formation of giant planets, and hence, the formation of planetary systems, we must be able to determine the interior structure and composition of giant planets. Jupiter and Saturn, our solar system's gas giants, combine to make up 92\% of the planetary mass of our solar system.  Giant planets are also vast natural laboratories for simple materials under high pressure in regimes that are not yet accessible to experiment.  With the recent rise in number and stunning diversity of giant planets, it is important to understand these planets as a class of astronomical objects.

We would like to understand the answers to basic questions about the structure and composition of these planets.  Are gas giants similar in composition to stars, predominantly hydrogen and helium with a small mass fraction of atoms more massive than helium of only $\sim 1$\%?  If these planets are enhanced in ``heavy elements'' (the $Z$ component) relative to stars, are these heavy elements predominantly mixed into the hydrogen-helium (H-He) envelope, or are they mainly found in a central core?  If a dense central core exists, how massive is it, what is its state (solid or liquid), and is it distinct or diluted into the above H-He envelope?  Can we understand if a planet's heavy element mass fraction depends on that of its parent star?  What methods of energy transport are at work in the interiors of these planets?  Does this differ between the gas giants and the ice giants?  Can we explain a planet's observable properties such as the luminosity and radius at a given age?

New data on the atmospheric composition or gravitational fields of our solar system's giant planets comes quite rarely, with long intervals between space missions that gather these precious data sets.  We are therefore at the mercy of both our own creativity, as we search for new ideas to explain the data we have, and at the mercy of technology, which allows us to push the boundaries of both experiment and computation.  The first decade of our new century is seeing a number of important advances in both the experiment and theory of materials at high pressure, so that we are in a better position to answer some of our questions outlined above.

Giant planets have long been of interest to physicists because they are natural laboratories of hydrogen and helium in the megabar to gigabar pressure range, at temperatures on the order of 10$^4$ K, which at the high pressure end is outside the realm of experiment.  The data that we use to shape our understanding of giant planets comes from a variety of sources.  Laboratory data on the equation of state (EOS, the pressure-density-temperature relation) of hydrogen, helium, ``ices'' such at water, ammonia, and methane, silicate rocks, and iron serve as the initial inputs into models.  Importantly, data are only avaible over a small range of phase space, so that detailed theoretical EOS calculations are critical to understanding the behavior of planetary materials at high pressure and temperature.  Within the solar system, spacecraft data on planetary gravitational fields allows us to place constraints on the interior density distribution for Jupiter, Saturn, Uranus, and Neptune. 

The year 1995 was Earth-shattering to the field of giant planets, as the first extrasolar giant planet 51 Peg b \cp{Mayor95} and also the first bona fide brown dwarf, Gliese 229B \cp{Nakajima95}, were discovered.  In particular the close-in orbit of 51 Peg b led to questions regarding its history, structure, and fate \cp{Guillot96,Lin96}.  Four years later, the first transiting planet, \hd\ \cp{Charb00,Henry00}, was found to have an inflated radius of $\sim$1.3 Jupiter radii (\rj), confirming that proximity to a parent star can have dramatic effects on planetary evolution \cp{Guillot96}.  However, the detections of nearly 50 additional transiting planets (as of May 2009) has raised more questions than it has answered.  For exoplanets, we often must make due with little information on planetary structure, namely a planet's mass and radius only.  For these planets, what we lack in detailed knowledge about particular planets, we can make up for in number.

Much further from their parent stars, young luminous gas giant planets are being directly imaged from the ground and from space \cp{Kalas08,Marois08}.  For these planets, planetary thermal emission is detected in a few bands, and a planet's mass determination rests entirely on comparisons with thermal evolution models, which aim to predict a planet's luminosity and spectrum with time.  As the initial conditions for planetary cooling are uncertain, the luminosity of young planets is not yet confidently understood \cp{Marley07,Chabrier07}.

In this paper we first discuss in some detail results of structural models of Jupiter, our ``standard'' example for gas giant planets.  We then look at similar models for Uranus and Neptune.  Our discussion then moves to calculations of the thermal evolution of our solar system's giant planets.  We then discuss current important issues in modeling exoplanets, and how these models compare to observations of transiting planets, as well as directly imaged planets.  We close with a look at the future science of extrasolar giant planets (EGPs).
 
\section{Core mass and metallicity of Jupiter, Uranus, \& Neptune}\label{NN}

\subsection{Introduction}\label{ssec:intro-NN}

In this section we address  the core mass and metallicity of Jupiter, Uranus and Neptune. In \S~\ref{ssec:Jup_McMZ} we compare results for Jupiter  obtained with different equations of state which are described in \S~\ref{ssec:EOS}. In \S~\ref{ssec:UN_McZZ} a large range of Uranus and Neptune structure models is presented that are consistent with the observed gravity data. Based on these models we discuss in \S~\ref{ssec:diss_core} the traditional concept of a rocky or icy core which is often used to derive implications for the formation process.

\subsection{EOS of H, He, and metals}\label{ssec:EOS}

\subsubsection{Matter inside the giant planets in the solar system}

Gas giant planets such as Jupiter and Saturn do not consist of \emph{gas} and icy giant planets such as Uranus and Neptune not of \emph{ice}. The gaseous phase of hydrogen, which is the predominant element of gas giant planets, becomes a non-ideal fluid at densities $\rho>0.01$ g/cm$^3$~\citep{SC95}. In Jupiter, this hydrogen density is reached in the outer 0.01\% of the total mass, and in Saturn in the outer 0.1\%. Similarly, the ice~{\small I} phase of water in Uranus and Neptune is left after only 0.02\% and the liquid phase after 0.2\% of the outer mass shell due to adiabatically rising temperature. The assumption of an adiabatic temperature gradient is important to the construction of state-of-the-art interior models \citep{SG04,Militzer08} and is supported by diverse observations. (See \S~\ref{ssec:method}.) This moderate rise of temperature accompanied with fast rising pressure towards deeper layers causes matter in giant planet interiors to transform to a warm, dense fluid, characterized by ionization, strong ion coupling and weak electron degeneracy. In Jupiter-size and Saturn-size planets, hydrogen, maybe helium too --depending on the EOS, metallizes giving rise to a strong magnetic field; in Neptune-size planets, water prefers (depending on the entropy) the ionically conducting superionic phase or the plasma phase~\citep{nettelmann09}.

Laboratory experiments for the EOS of warm dense matter are very challenging. To date, the EOS of H is well constrained below $\sim$0.3~g cm$^{-3}$ and below $\sim$25~GPa (0.25 Mbar) by precise gas gun shock compression experiments. See \citet{SG04} for an overview of data from compression experiments of deuterium and \citet{French09} for water. At larger densities and pressures however, as relevant for planetary interiors, experimental data have large error bars and single-shock data (Hugoniots) bend towards higher temperature regions in the phase diagram than are relevant for solar system giant planets. High-precision multi-shock experiments are urgently required to constrains the EOS of hydrogen. Until then, there is much space for theoretical EOS. Next we will describe seven EOS that are consistent with experimental EOS data, and have been applied to Jupiter models as well. Five of them are based on the chemical picture of distinct species interacting via specific effective pair potentials, and the other two are based on the physical picture of electrons and nuclei interacting via Coulomb forces \cp[see, e.g.][]{SC95}.

\subsubsection{Hydrogen EOS}

\paragraph{\textbf{Sesame:}} The H-EOS \emph{Sesame 5251} is the deuterium EOS 5263 scaled in density as developed by~\citet{Kerley1972}. It is built on the assumption of three phases: a molecular solid phase, an atomic solid phase, and a fluid phase that takes into account chemical equilibrium between molecules and atoms and ionization equilibrium of the fluid phase of atoms. A completely revised version  by~\citet{Kerley2003} includes, among many other improvements, fits to more recent shock compression data resulting into larger~(smaller) compressibility at $\sim$0.5~(10)~Mbar. In this article we call this improved version \emph{H-Sesame-K03}. \citet{SG04} patched the original version at pressures between 100 bar and 0.4~Mbar with another EOS in order to reproduce the gas gun data and call this version \emph{H-Sesame-p}.

\paragraph{\textbf{H-SCVH-i:}} This widely used EOS omits the astrophysically irrelevant region of cold dense solid hydrogen and relies on the free-energy-minimization technique throughout the $\rho-T$ region that is relevant for giant planets and low-mass stars. As in the fluid phase of Sesame~5251, it takes into account the species H$_2$, H, H$^+$, and $e$. But at the transition to metallic hydrogen, thermodynamic instabilities are found and considered as a first-order phase transition, the \emph{Plasma Phase Transition}. In an alternative version, \emph{H-SCVH-i}, the instabilities are smoothed out by careful interpolation between the molecular and the metallic phase. Details are given in~\citet{SC95}.

\paragraph{\textbf{LM-SOCP and LM-H4:}} These EOS are modifications of the simple linear mixing model of Ross (see e.g.~\cite{Holmes95}). It assumes the total Helmholtz free energy $F$ of a system of H$_2$ molecules and metallic H as linear superposition of the single components' free energies $F_{\rm mol}$ and $F_{\rm met}$, respectively. The original EOS was constructed to fit the gas gun data by adjusting the effective molecular pair potential, and to fit their low reshock temperatures by addition of a fitting term $F_{\rm fit}$ in the total free energy. This term causes a region where $\nabla_{ad}<0$ along the Jupiter isentrope. \citet{SG04} avoided this behavior by taking into account electron screening in the metallic component (LM-SOCP) or by the admixing of $D_4$ chains as an additional species (LM-H4).

\paragraph{\textbf{DFT-MD:}} Applying density-functional molecular dynamics (FVT-MD) to simultaneous simulation of H and He nuclei (100 H and 9 He nuclei in periodic boundary conditions), \citet{Militzer08} were the first to provide an EOS including H/He mixing effects for a broad range of densities $\rho$ and temperatures $T$ relevant for Jupiter's interior. They used the CPMD code with Troullier-Martins norm-conserving pseudopotentials and the VASP code with projector augmented wave pseudopotentials to generate EOS data at $\rho\geq 0.2$~g/cm$^3$ and $T\geq 500$~K and used classical Monte Carlo simulations at smaller densities. H/He mass mixing ratios other than 0.2466 were realized by diminishing the density along the J-isentrope in accordance with He EOS data. 

\paragraph{\textbf{H-REOS:}} For hydrogen densities $0.2\leq\rho\leq 9$~g/cm$^3$ and temperatures $1000\leq T\leq 30000$~ K, \cite{nettelmann08} also use the VASP code developed by \citet{KH93a,KH93b} and \citet{KF96}. The main differences in the calculation of EOS data for H/He mixtures compared to DFT-MD EOS by \cite{Militzer08} are i) inclusion of finite temperature effects on the electronic subsystem by Fermi weighting of the occupation of bands before minimizing the electronic energy density functional (FT-DFT), (ii) separate FT-DFT molecular dynamics calculations for H and He with subsequent linear mixing, iii) application of FVT$^+$ ~\citep{Holst07} to generate H EOS data at lower densities and higher temperatures. FVT$^+$ combines fluid variational theory, a minimization method for the free energy of neutral hydrogen, with Pade formulas for fully ionized hydrogen taking into account ionization equilibrium. While FVT$^+$ predicts a plasma phase transition between 0.27 and 0.5~g/cm$^3$, H-REOS does not, since it transitions smoothly from FVT$^+$ to FT-DFT-MD data below 0.2~g/cm$^{3}$.

\paragraph{\textbf{Other EOS:}} It is interesting to note that there are H-EOS that do not give  acceptable Jupiter models, indicating an invalid $\rho-P$ relation at those pressures where the Jovian gravity field is most sensitive to the internal mass distribution. Among such EOS are LM-B~\citep{SG04} and FVT$^+$. In this sense, Jupiter interior models serve as a check of EOS data in the warm dense matter regime.

\subsubsection{He EOS}

Helium equations of state used together with the hydrogen equations of state described above are listed in Tab.~\ref{tab:J-EOS}. The He EOS \emph{He-SCVH} is described in~\citet{SC95}, \emph{He-Sesame-K04} in~\citet{Kerley2004b}, \emph{He-REOS} in~\citet{nettelmann08}, and \emph{DFT-MD} in~\citet{Vo07}, respectively.  Relative differences in pressure and internal energy along relevant isotherms are within $\sim$~30\%, comparable to those of the H EOSs. With an average H/He particle number ratio  below 1/10, the effect of the He EOS on giant planet interior models lies less in its $\rho-P$ relation but more in its mixing ability with hydrogen and the possibility of He sedimentation. This topic is addressed in \S 2.6.

\subsubsection{EOS of metals}

Diverse EOS of heavy elements are used to represent \emph{metals} within Jupiter's envelope and core. \citet{SG04} take the Sesame EOS 7154 of water to represent ices\footnote{The label \textit{ice} refers to a mixture of H$_2$O, CH$_4$, and NH$_3$ that are supposed to have been in an ice phase during protoplanetary core formation.} (I), and the Sesame EOS 7100 of dry sand to represent rocks (R) with an upper limit of the R-component of 4\% in the envelope. \cite{nettelmann08} either scale He-REOS in density by a factor of four \emph{(He4)} or use \emph{H2O}-REOS. This new EOS of water is a combination of accurate ice~{\small I} and liquid water data, FT-DFT-MD data at densities and temperatures relevant for giant planet interiors, and Sesame 7150 at small densities and high temperatures with interpolated regions in between to smoothly join these different data sets. They assume a rocky core using the fit-formula to experimental rock data below 2 Mbar by \cite{Hubbard89}. Rocks lead to roughly 50\% less massive cores than ices. \citet{Kerley2004a} represents the core material by SiO$_2$. For metals in the envelope, he assumes an initial composition of O, C, N, and S of relative solar abundance in the outer region with the addition of Si and Fe in the inner region of Jupiter. For a given enrichment factor, the chemical equilibrium abundances of molecules formed by these species in a H/He mixture is calculated and the corresponding EOS tables of the occurring components are added linearly to the H/He EOS.

\subsection{Construction of interior models: constraints and methods}\label{ssec:method}

\paragraph{Constraints.} For interior models of the solar system giant planets, in general the following observational constraints are taken into account: the total mass $M$, the equatorial radius $R_{eq}$, the 1-bar temperature $T_1$, the angular velocity $\omega$, the gravitational moments $J_{2n}$, in particular $J_2$ and $J_4$, the atmospheric He mass fraction $Y_1$, and occasionally the atmospheric abundances of volatile species, except oxygen. Due to low atmospheric temperatures, O, if present, is believed to condense out as H$_2$O clouds at higher pressures deeper inside the planets. These pressures have not yet been reached by observation, such that an observed O abundance is believed to not be indicative for the overall abundance in the envelope. On the other hand, the measured supersolar abundances of other volatiles are generally explained by the dissolution of volatile-rich icy planetesimals that were captured by the young forming planet, implying a supersolar overall water abundance. In the absence of representative data, the O abundance is usually assumed of the order of other volatiles abundances~\citep{Kerley2004a}. The mean He content, $\bar{Y}$, cannot be observed, but from solar evolution theory in accordance with observational data for the sun, a value of $Y=0.275\pm 0.01$ is generally accepted as a constraint for planet interior models~\citep{FH03}. Beside the uncertainties in the equation of state, the error bars of the observables give rise to broad sets of models for a single planet.

\paragraph{Methods.} The luminosity is an important observable for evolution models, as described below in \S~\ref{JF:sec:evolJS}.  For structure models, it is important in the sense that it gives a hint of the temperature profile. The high intrinsic luminosities of Jupiter, Saturn, and Neptune for instance strongly point towards an adiabatic, convective interior on large scales, since energy transport by radiation or conduction are too ineffient to account for the observed heat flux \cp{Hubbard68}.  This is because of frequent collisions in the dense interior and strong molecular absorption in the less dense outer region. Convection, which will tend to homogenize the planet, leads to an adiabatic temperature gradient. In the absence of a convection barrier, the envelope of a giant planet can be assumed adiabatic (isentropic) and homogeneous, where the entropy is fixed by $T_1$ \citep{Hubbard73}.

Given $M_p$, to reproduce $R_p$ one has to either make the additional assumption of a core of heavy elements, or to choose a particular envelope metallicity $Z$, since Jupiter and Saturn are smaller in radius than pure H-He planets \cp[e.g.][]{DeMarcus58,Podolak74}. Thus the radius fixes the core mass $M_{core}$ or, alternatively, $Z$. This property is used to derive a core mass or metallilicity of transiting extrasolar planets, since only the mass and radius can be measured. Furthermore, the \emph{Voyager} and \emph{Galileo probe} measurements give $Y_1<\bar{Y}$ for Jupiter and Saturn, implying either an inhomogeneous interior, or $\bar{Y}$ below the cosmological value, or a mixing barrier dividing the interior into a He-depleted outer envelope with $Y=Y_1$ and a He-enriched inner envelope. Most modelers prefer the last scenario. There are several possibilities for where to locate the layer boundary, characterized by the transition pressure $P_{12}$ between the outer (layer 1) and inner (layer 2) envelope, depending on the mechanism causing the He discontinuity. Candidates are a first-order phase transition, e.g. a plasma phase transition of H whose existence is still a matter of debate, and H/He phase separation with He sedimentation. For practical  purposes, $P_{12}$ can be varied within a reasonable range around 3~Mbar. For Uranus and Neptune, $Y_1$ is consistent with $\bar{Y}$ within the observational error bars.

While $\omega$ enters the equations to be solved explicitly, the gravitational moments $J_2$ and $J_4$ have to be adjusted within an iterative procedure and thus require two further free parameters. These can be the metallicities $Z_1$ and $Z_2$ in the two envelope layers. More generally, the parameters $Z_1$, $Z_2$, and $M_{core}$ are used to adjust $J_2$, $J_4$, and $R_{eq}$ \citep{Chabrier92,Guillot99,nettelmann08}. Other authors do not allow for a discontinuity of metals~\citep{SG04,Militzer08}. An argument in favor of $Z_1=Z_2$ is large-scale convection of the hot, young planet; an argument in favor of $Z_1\not=Z_2$ is core-accretion formation with inhomogeneous planetesimal delivery in the envelope leading to early formation of a convection barrier, due to mean molecular weight gradients.  However, if remnant planetesimal gradients are present, it is unlikely that they could be characterized simply by one number, $Z_2$.  Furthermore composition gradients inhibiting convection would void the assumption of an adiabatic interior.

Table \ref{tab:J-EOS} gives an overview about the Jupiter model series and the EOS used therein, the underlying different structure type assumptions (discontinuities in $Y$ and $Z$). We present and discuss results for Jupiter's core mass and heavy element abundance in the following subsection.

\begin{table}[hbt]
\renewcommand{\arraystretch}{1.8}
\caption{Overview of Jupiter model series.}\label{tab:J-EOS}
\begin{tabular}{llllll}
\hline\noalign{\smallskip}
Name				& H-EOS				& He-EOS			& Z-EOS					& type							& Ref. \\
(EOS)			&						&						&							&									& (J)	 \\
\hline\noalign{\smallskip}
SCVH-I-99		& H-SCVH-I		& He-SCVH			& He-SCvH				& \parbox[t]{1.5cm}{
																					 	  		$Y_1<Y_2$\\
																						 		$Z_1\neq Z_2$}				& (1)  \\
SCVH-I-04		& H-SCVH-I		& 	He-SCVH			& \parbox[t]{2cm}{
																	Sesame 7154,\\ 	
                           						Sesame 7100}		& \parbox[t]{1.5cm}{  	
																						  		$Y_1<Y_2$\\	
																						  		$Z_1=Z_2$}					& (2) \\
LM-SOCP			& LM-SOCP			& ''					& ''						& ''								& ''\\
LM-H4			& LM-H4				& ''					& ''						& ''								& ''\\
Sesame-p		& H-Sesame-p	& ''					& ''						& ''								& ''\\
Sesame-K04	& Sesame-K03	& Sesame-K04	& \parbox[t]{2cm}{
																	linear mixture of H$_2$O, CH$_4$, NH$_3$, 
																	C, N, O, H$_2$S, S, SiO$_2$, Fe}	
																								& \parbox[t]{1.5cm}{
																								$Y_1<Y_2$\\
																								$Z_1<Z_2$}				& (3)\\	
LM-REOS			& H-REOS			& He-REOS			& \parbox[t]{2cm}{
																	H$_2$O-REOS,\\
																	He4-REOS}				& \parbox[t]{1.5cm}{
																								$Y_1<Y_2$\\
																								$Z_1<Z_2$}					& (4)\\
DFT-MD			& DFT-MD			& DFT-MD			& CH$_4$, H$_2$O	& \parbox[t]{1.5cm}{
																								$Y_1=Y_2$\\
																								$Z_1=Z_2$}					& (5)\\
\noalign{\smallskip}\hline\noalign{\smallskip}
\end{tabular}
\begin{minipage}{\textwidth}\small
References for Jupiter models: (1)~\cite{Guillot99}, (2)~\cite{SG04}, (3)~\cite{Kerley2004a}, (4)~\cite{nettelmann08}, (5)~\cite{Militzer08}. In all cases: $Y_1=Y_{\rm atm}=0.238$.
\end{minipage}
\end{table}

\subsection{Results: Core mass and metallicity of Jupiter}\label{ssec:Jup_McMZ}

\begin{figure}[h]
\includegraphics[angle=0,width=\textwidth]{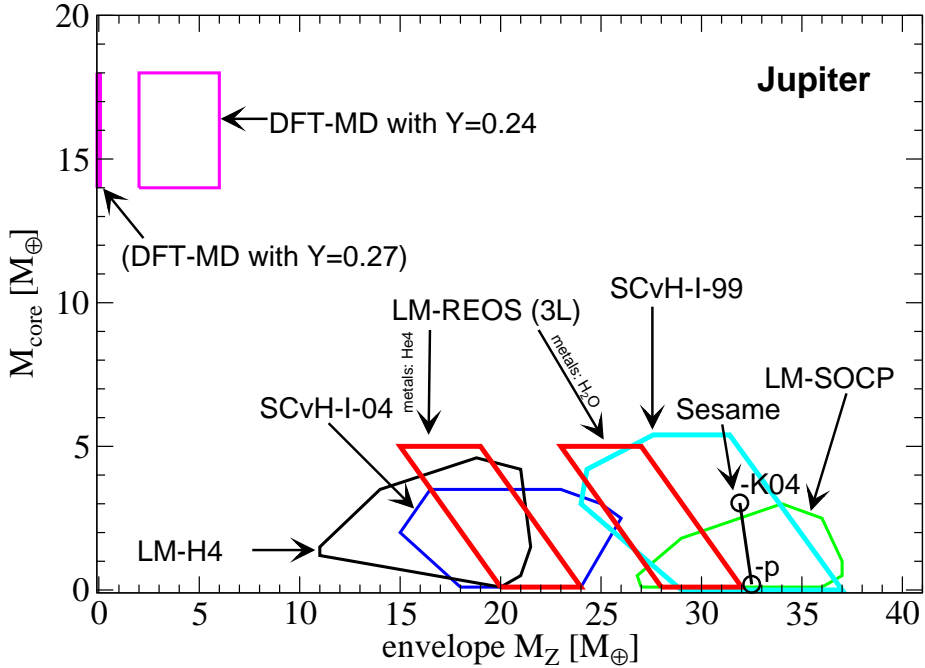}
\caption{(Color online) Mass of the core and of heavy elements within the envelope. Each box represents the solutions found using a particular equation of state as listed in Tab.~1. In the case of DFT-MD EOS models which originally have $\bar{Y}=0.238$, we also indicate the position if 3\% ($\sim 9M_{\oplus}$) of metals are replaced by He in order to have $Y=0.27$, in accordance with all other models in this figure.}
\label{fig:McMZ} 
\end{figure}

Figure~\ref{fig:McMZ} shows the resulting mass of the core and the mass $M_Z$ of metals in the envelope(s) found by different authors using the diverse EOS as listed in Tab.~\ref{tab:J-EOS}. Note that all these solutions have $\bar{Y}=0.275\pm 0.01$ except DFT-MD models, which have $\bar{Y}=0.238$. To better compare these solutions, enhancing $\bar{Y}$ by 0.03 to 0.27 in the latter solutions requires replacing $\sim 9M_{\oplus}$ of metals by He. In this case, DFT-MD models have metal-free envelopes. To avoid this problem, \citet{Militzer08} suggest a He layer above the core due to He sedimentation yielding rocky core masses of 5-9$M_{\oplus}$, instead of 14-18 $M_{\oplus}$, in better agreement with all other solutions. The other extreme of high envelope metallicity, up to $37M_{\oplus}$, is found by using LM-REOS or SCvH-I-99. To show the effect of the EOS of metals, models using the He EOS scaled in density by a factor of 4 (He4) and using water for metals are presented. Heavier elements, i.e. magnesium-silicates, would give even lower $M_Z$ values. We conclude from this figure that the choice of composition and EOS for the metals has a large effect on the envelope metallicity and a small effect on the core mass. If these EOS reflect our current knowledge, we conclude that the interior of Jupiter is badly constrained with a possible core mass ranging from 0 to 18 \me\ and an envelope heavy element ($Z$) mass from 0 to 37$M_{\oplus}$. If these large uncertainties are taken at face value, a prediction about Jupiter's formation process is highly unreliable.

\begin{figure}[h]
\includegraphics[angle=0,width=\textwidth]{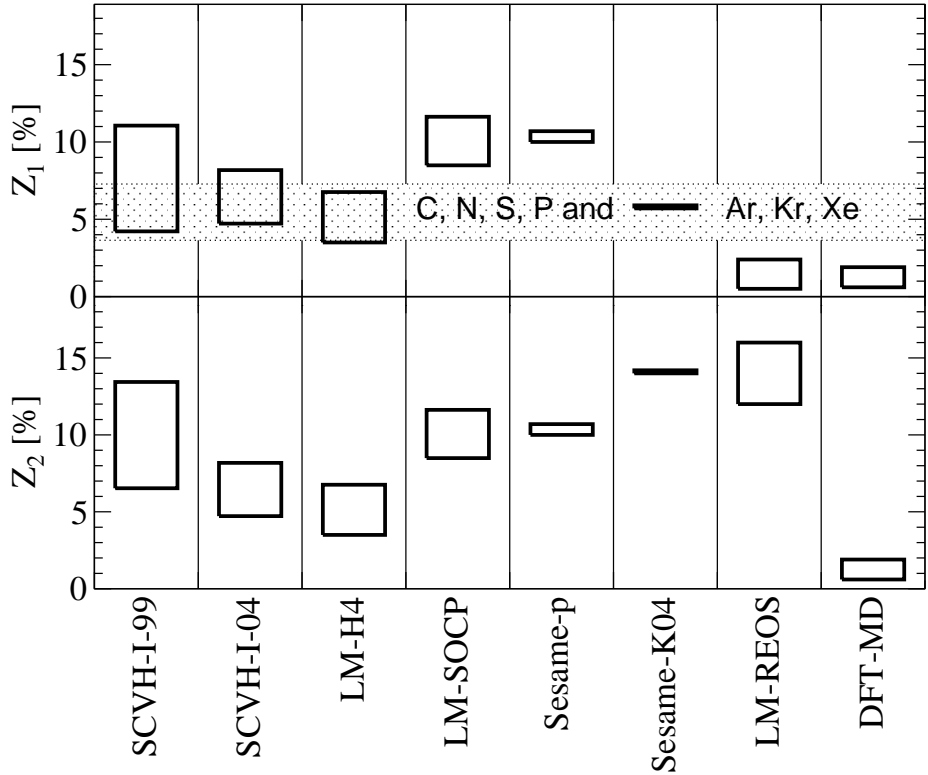}
\caption{Mass fraction of heavy elements in the outer envelope ($Z_1$) and the inner envelope ($Z_2$) of Jupiter interior models using the different equations of state described in \S~\ref{ssec:EOS}. The dotted region shows the atpmosheric metallicity \emph{if} the O abundance is similar to the values measured for C, N, S and some noble gases, i.e. $2-4$ times solar.}
\label{fig:ZZ} 
\end{figure}

Figure~\ref{fig:ZZ} shows the mass fraction of metals in the two envelopes for the same EOSs as in Fig.~\ref{fig:McMZ}. Models without a discontinuity of metals have $Z_1=Z_2$ per definition. For tentative evaluation of these results, $Z_1$ is compared with the range of atmospheric abundances of some volatile species, where we used two assumptions. The first is that O atoms are as abundant as the species C, N, S and Ar, Kr, Xe, i.e. $2-4$~$\times$ solar~\citep{Mahaffy00}, and the second is a mass fraction equivalent of $1\times$ solar $\simeq$1.9\%. As stated in \S~\ref{ssec:method}, the real O abundance $\rm x_O$ in Jupiter might be much higher than the measured value of 30\% of the solar value due to condensation of water above 20~bar, where the \emph{Galileo} probe stopped working~\citep{Wong04}. If however $\rm x_O\ll x_{C,N,S,P}$, then the lower boundary of the dotted region in Fig.~\ref{fig:ZZ} would sink, otherwise if $\rm x_O\gg x_{\rm C,N,S,P}$, then the upper boundary would rise. 

For interior modeling, there are several assumptions that affect the resulting envelope metallicity. All EOS except DFT-MD use the simplifying linear mixing approximation to combine H- and He-EOS. In case of the DFT-MD EOS however, which takes into account the mixing effect by simultaneous simulation of H and He atoms, an up to 5\% volume enhancement (density decrement) is found compared to linear mixing~\citep{Vo07} at pressures and temperatures typical for Jupiter's deep outer envelope, where $J_4$ is most sensitive to the metallicity. Compensating for this reduction in density of the H/He subsystem requires a corresponding enhancement in metals. Thus the calculated $Z_1$ values might increase by up to 5 percentage points, except for DFT-MD EOS. Furthermore, Jupiter's cloud patterns are known to rotate on cylinders with different velocities as a function of latitude. If differential rotation extends into the interior, the gravitational moments calculated by assuming rigid body rotation have to be corrected. ~\cite{ZT78} suggest a small correction of 0.5\% for $J_2$ and 1\% for $J_4$ based on observations of atmospheric winds \cp[see also][]{Hubbard82}; \citet{JLiu08} predict a penetration depth of deep-zonal winds down only 0.04$R_{\rm J}$, supporting an only slight effect on the low-degree gravitational harmonics. \cite{Militzer08} on the other hand invoke interior winds penetrating 10\% into Jupiter's envelope in order to match $J_4$, which otherwise would differ from the observed value by more than two standard deviations. Applying the same correction necessary for DFT-MD models on interior models using LM-REOS, which exhibits the smallest $Z_1/Z_2$ ratio (see Fig.~\ref{fig:ZZ}), gives $Z_1/Z_2>1$. 

We conclude that future spacecraft-based measurements are desirable in order to constrain the envelope metallicity and, consequently, narrow the set of H/He equations of state currently offered. Among the most helpful observations we suggest a measurement of Jupiter's O abundance at pressures between 10 and 60 bar (above and below the liquid water to vapor transition along the isentrope), and a determination of deep-zonal winds by measuring high-order harmonics.  NASA's forthcoming \emph{Juno Mission} will indeed measure these harmonics, as well as constrain the deep water and ammonia abundances from microwave spectra \cp{Matousek07}.

\subsection{Results: Core mass and metallicity of Uranus and Neptune}\label{ssec:UN_McZZ}

\begin{figure}[hbt]
\includegraphics[angle=0,width=\textwidth]{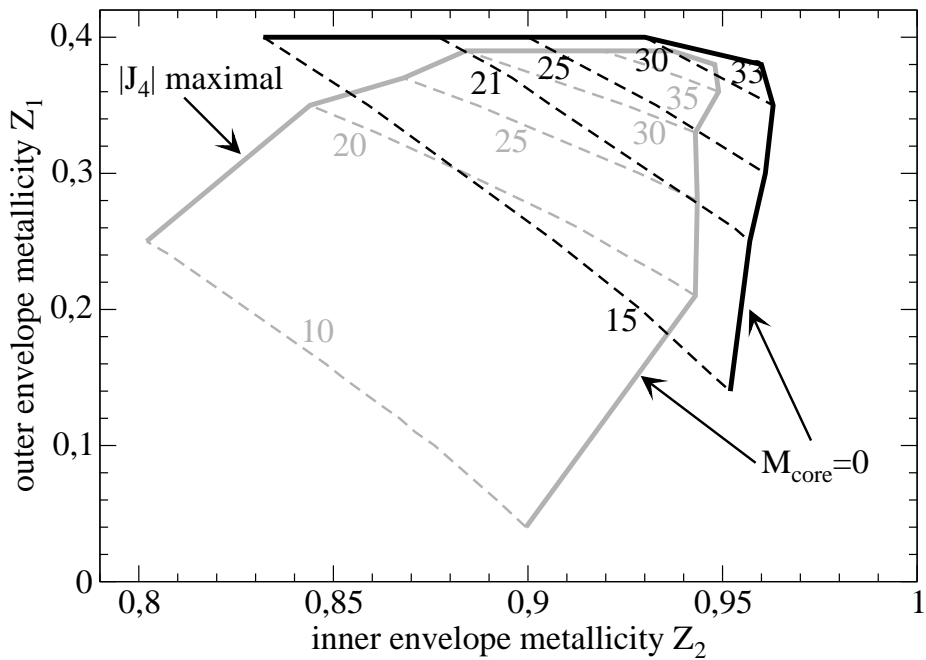}
\caption{Mass fraction of metals in the outer envelope ($Z_1$) and in the inner envelope ($Z_2$) of three-layer models of Uranus (grey) and Neptune (black). The thick solid lines indicate the range where solutions have been found. Numbers at dashed lines give the transition pressure $P_{12}$ in GPa, and dashed lines show the behavior of solutions if $P_{12}$ is kept constant and $J_4$ is varied within the $1\sigma$-error. Increasing $Z_1$ increases $|J_4|$. No Uranus solutions are found above the upper thick line. Neptune's $J_4$-error bar is large, so we stopped arbitrarily at $Z_1$=40\%. Decreasing $Z_1$ results into higher $Z_2$ values and smaller core masses. Below the lower thick lines, no solutions exists. These models are based on LM-REOS using water for metals.}
\label{fig:UN} 
\end{figure}

We apply the same method used for Jupiter interior calculations with LM-REOS for three-layer models of Uranus and Neptune. Planet models consist of a two-layer envelope and core.  Envelope metals are represented by water and the core consists of rocks. Uranus and Neptune have large observational error bars of $J_4$ of 10\% and 100\%, respectively.  Results are shown in Fig.~\ref{fig:UN}.  For a given transition pressure $P_{12}$ between the two envelopes composed of mixtures of H/He and water (layer 1 and 2), the solutions move along almost straight lines, and changing $P_{12}$ causes a parallel shift of the line.  Decreasing $Z_1$ requires a higher inner envelope metallicity ($Z_2$) in order to match $J_2$. Simultaneously, the mass of the core (layer 3) shrinks with $M_{core}=0$ defining the maximal possible $Z_2$ value for a given layer boundary.  Here, pure water envelopes are not allowed. Replacing a H/He mass fraction of 5\%~(10\%, 12\%) by the molecular weight of CH$_4$ results in a H$_2$O/CH$_4$ mass ratio of 0.6~(0.2, 0), but those models with an inner envelope of pure 'icy' composition have not been calculated here. On the other hand, replacing some H$_2$O by rocks will result into a higher H/He fraction, and in the more realistic case of a solar ice/rock ratio of $\sim$2.7, H/He free deep envelopes are not possible. These results are in good agreement with those by~\cite{Hubbard89}.

Most Uranus and all Neptune models presented here have also a significant heavy element (water) enrichment in the outer envelope ($P<P_{12}$). An upper limit of Uranus' $Z_1$ is given by the requirement to meet $J_4$; for Neptune, the large error bar of $J_4$ allows for even higher outer envelope metallicities than 0.4. In any case, all models have a pronounced heavy element discontinuity. No Uranus (Neptune) models are found with $P_{12}>$38 (33)~GPa because of $M_{core}\to 0$. We did not calculate models with $P_{12}<10$~GPa, since this discontinuity is perhaps caused by the transition from molecular water to ionic dissociated water, which occurs around 20~GPa~\citep{French09}.

Uranus and Neptune are very similar planets with respect to their core mass and total heavy element enrichments~\citep{Nbook}, and are very different planets with respect to their internal heat fluxes, as well as to observed molecular species. While C in both planets is about 30-60 times solar, CO and HCN have been detected in Neptune, but not in Uranus~\citep{Gautier95}, likely indicating the absence of efficient convective transport in Uranus. Convection can be inhibited by a steep compositional gradient or by a region with sufficiently high conductivity. Calculations by~\cite{Guillot94a} suggest the presence of such a radiative region at 1000 K in Uranus\footnote{These calculations should be revisited in light of since-discovered strong opacity sources in the deep atmosphere of Jupiter, which close its previously postulated radiative window \cp{Guillot04}}.  Since the temperature in a radiative layer rises less than in the adiabatic case, this explanation for Uranus' small heat flux tends to smaller present-day central temperatures. At layer boundaries induced by a steep compositional gradient on the other hand, the temperature rises faster than in the adiabatic case leading to higher present-day central temperatures. One step forward to decide as to the more appropriate scenario could be a calculation of cooling curves using non-adiabatic temperature gradients. The  good agreement of Neptune cooling curves based on two adiabatic, homogeneous layers of pure H/He and water~(see below, and M. Ikoma, \textit{personal communication} 2008) with the present luminosity possibly shows us that Neptune's structure may not necessarily be extremely complex.

\subsection{Results: Evolution of Jupiter and Saturn } \label{JF:sec:evolJS}
\begin{figure}[hbt]
\includegraphics[angle=0,width=\textwidth]{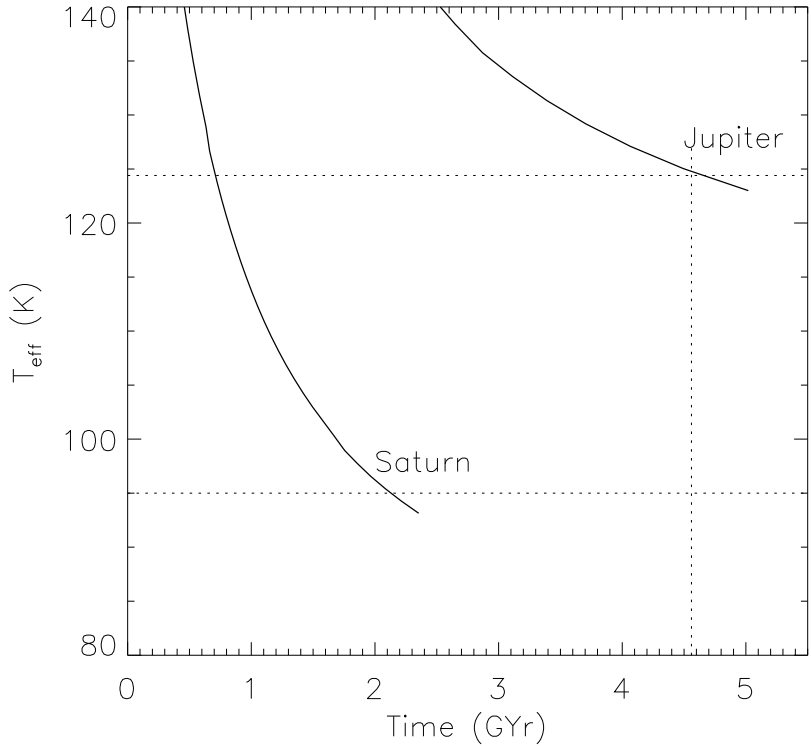}
\caption{Homogeneous evolutionary models of Jupiter and Saturn, adapted from \ct{FH03}.  The solar system's age as well as the \teff\ of Jupiter and Saturn are shown with dotted lines.}
\label{js}
\end{figure}

Our understanding of the evolution of Jupiter and Saturn is currently imperfect. The most striking discrepancy between theory and reality is Saturn's luminosity. Saturn's current luminosity is over 50\% greater than one predicts using a homogeneous evolution model, with the internally isentropic planet radiating over time both its internal energy and thermalized solar radiation. This discrepancy has long been noted \cp{Pollack77,Grossman80,Guillot95,Hubbard99}.  Homogeneous evolutionary models of Saturn tend to reach an effective temperature of 95.0 K (Saturn's current known \teff) in only 2.0-2.7 Gyr, depending on the hydrogen-helium equation of state (EOS) and atmosphere models used. However, purely homogeneous models appear to work well for Jupiter. Figure \ref{js} shows homogeneous evolutionary models for both planets from \ct{FH03}. It has also long been believed that the most promising route to resolving this discrepancy is the possible phase separation of neutral helium from liquid metallic hydrogen in the planet's interior, beginning when Saturn's effective temperature reached 100-120 K \cp{SS77a,SS77b}.  This sinking of ``helium rain" can be an appreciable energy source.

\begin{figure}[hbt]
\includegraphics[angle=0,width=\textwidth]{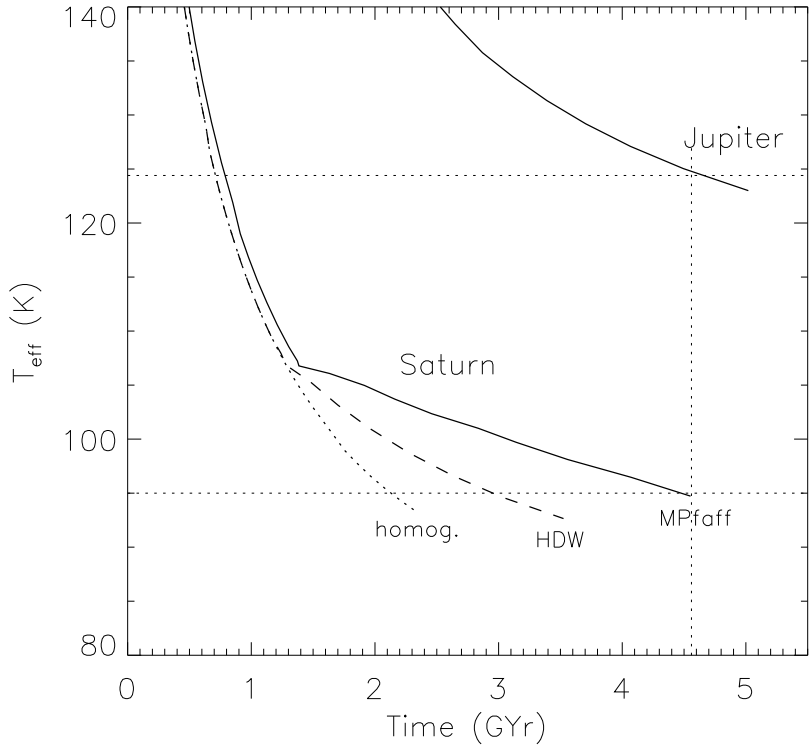}
\caption{Evolutionary models of Saturn including helium phase separation, adapted from \ct{FH04}.  ``HDW'' uses the H/He phase diagram of \ct{HDW}, which allows immiscible helium to redissolve at higher pressures and hotter temperatures in the liquid metallic hydrogen.  ``MPfaff'' is an ad-hoc phase diagram that forces immiscible helium to rain down to Saturn's core.}
\label{js2}
\end{figure}

\ct{FH03} tested a variety of high-pressure H/He phase diagrams that had been published since the mid 1970's.  Of particular note, they found that the phase diagram of \ct{HDW}, which is essentially the same as that of \ct{Stevenson75}, is inapplicable to the interiors of Jupiter and Saturn, if helium phase separation is Saturn's only additional energy source. As Figure \ref{js2} shows, this phase diagram prolongs Saturn's cooling only 0.8 Gyr, even in the most favorable circumstance that all energy liberated is available to be radiated, and does not instead go into heating the planet's deep interior. 

\ct{FH03} next inverted the problem to derive an ad-hoc phase diagram that could simultaneously explain Saturn's current luminosity as well as its current atmospheric helium abundance \cp{CG00}.  The helium abundance is depleted relative to the Sun, and is consistent with helium being lost to deeper regions of liquid metallic hydrogen at Mbar pressures.  The ad-hoc phase diagram forced helium that rained out to fall all the way down to Saturn's core, thereby liberating a significant amount of gravitational potential energy.  In light of the new first principles calculations of H/He phase diagrams \cp{Lorenzen09, Morales09}, thermal evolution models of Jupiter and Saturn should now be revisited.

\subsection{Results: Evolution of Uranus and Neptune} \label{NN:sec:evolUN}

In the previous section we have seen that homogeneous evolution models work well for Jupiter, but not for Saturn , yielding cooling times that are too short. In this section we will see that homogeneous evolution models work fairly well for Neptune, but certainly not for Uranus, yielding cooling times too long to be consistent with the age of the solar system. 

The general results of \S 2.6 and 2.7 is based on solving the common energy balance equation
\begin{equation}\label{eq:lumis}
L-L_{\odot} = L_{int}
\end{equation}
where $L(t)=4\pi R^2(t)\sigma T_{\rm eff}(t)^4$ is the luminosity (mostly measured as flux in the mid infrared) of the planet attributed to an effective temperature $T_{\rm eff}$. Here, $L_{\odot}(t)=4\pi R^2(t)\sigma T_{\rm eq}(t)^4$ is the luminosity due to only to thermalized and reradiated absorbed solar flux, as parameterized by the equilibrium temperature $T_{eq}$, the \teff\ that planet would have in case of no intrinsic luminosity, $L_{int}(t)$. Taking into account cooling and ongoing gravitational contraction as energy sources to supply the radiative losses, we can write
\begin{equation}\label{eq:lumiint}
L_{int}(t) = - \int_0^M dm\, T(m,t)\frac{\partial s(m,t)}{\partial t}
\quad,
\end{equation}
where $T(m,t)$ is the internal temperature profile at time $t$ and $s(m,t)$ is the specific entropy. With a relation between the \teff\ and the atmospheric temperature at say, 1 bar \cp[see, e.g.][for detailed atmosphere models for warmer planets]{Burrows97},  Eqs.~\ref{eq:lumis} and \ref{eq:lumiint} can be converted into a single differential equation for $T_{\rm eff}(t)$.  Often an arbitrary initial condition is used (see \S \ref{youngj}) and the early \teff\ drops very quickly, such that planets ``forget'' their initial conditions.  Other cooling curves can be obtained only by introducing an additional free parameter or by assuming a birth that is colder than the arbitrarily hot start.

In an earlier investigation of Uranus and Neptune cooling models, \citet{Nbook} for instance assume mean values for the internal temperature and the specific heat $c_v=Tds/dT$ of the planetary material, neglect the relatively small contributions to the intrinsic luminosity from current gravitational contraction, and allowed a cold start. Alternatively, they introduce a variable fraction of the thermal heat content that contributes to the intrinsic luminosity, i.e. that lies within the convectively unstable, homogeneous region of the planet. Based on these assumptions they find that both Uranus and Neptune's cooling time would exceed the age of the Solar system with a larger deviation of some gigayears for Uranus, necessitating either a cold start or a significant fraction of the interior does not contribute to $L_{int}$.

\begin{figure}[hbt]
\includegraphics[angle=270,width=\textwidth]{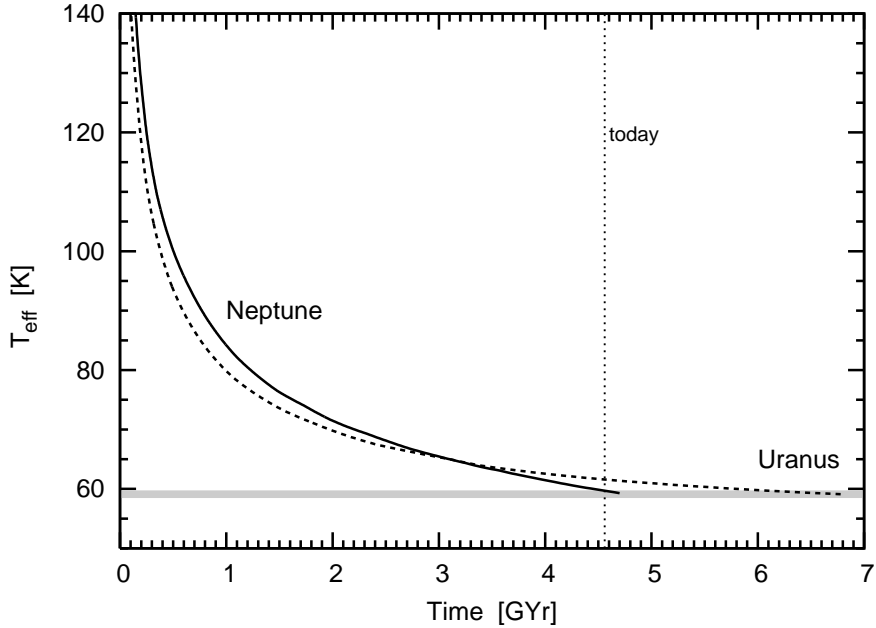}
\caption{Homogeneous evolutionary models of Uranus (dashed) and Neptune (solid). The underlying interior models are among those presented in Fig.~\ref{fig:UN}.  Notably, the real Uranus is underluminous as compared to the model.  The solar system's age is shown (dotted line) and the grey bar indicates the present $T_{\rm eff}$s.}
\label{fig:UN_evol}
\end{figure}

Qualitatively the same result of $\Delta t_{\rm cool}^{(\rm U)} \gg \Delta t_{\rm cool}^{(\rm N)} > 4.56$~GYr was reported by M. Ikoma (\textit{personal communication} 2008) for fully differentiated models with three homogeneous layers (rock core, ice layer, H/He envelope) using diverse ice equations of state.
However, such a centrally condensed interior structure is not consistent with the gravity field data of Uranus and Neptune, as discussed above. In this work we present in Fig.~\ref{fig:UN_evol} evolution tracks based on interior models within the sets of acceptable present-day solutions from Fig.~\ref{fig:UN}. The Uranus model in Fig.~\ref{fig:UN_evol} has $P_{12}=25$~GPa, $Z_1=0.35$, $Z_2=0.887$, $M_{core}=1.48M_{\oplus}$, and the Neptune model has $P_{12}=21$~GPa, $Z_1=0.37$, $Z_2=0.896$, $M_{core}=1.81M_{\oplus}$. The thick grey line indicates the uncertainty of their present day $T_{\rm eff}$ of 59.1 and 59.3~K, respectively. Note that while $T_{\rm eff}^{\rm (U)}\simeq T_{\rm eff}^{(\rm N)}$ and $R^{(\rm U)}\simeq R^{(\rm N)}$, we have $L^{(\rm U)}\simeq L^{\rm (N)}$, but $L_{int}^{\rm (U)} < L_{int}^{\rm (N)}$ because of Eq.~\ref{eq:lumis} and $L_{\odot}^{(\rm U)} > L_{\odot}^{(\rm N)}$ (as Neptune is less irradiated). With the same underlying relation between effective and atmospheric temperature and the same equation of state (LM-REOS), homogeneous cooling of Neptune gives roughly an age of 4.6~Gyr, but for Uranus of $\approx 2.5$~Gyr more.

Obviously, the results $\Delta t_{\rm cool}^{(\rm U)} \gg 4.56$~Gyr and $\Delta t_{\rm cool}^{(\rm N)} > 4.56$~Gyr appear insensitive to the details of the structure model and of the equations of state used. Hence we must call into question the assumption of  convective envelope(s) beneath all of these models.

Convection can be inhibited by a steep compositional gradient or by a region with sufficiently high conductivity. Calculations by~\citet{Guillot94a} suggest the presence of such a radiative region at 1000 K in Uranus.  Since the temperature in a radiative layer rises less than in the adiabatic case, this explanation for Uranus' small heat flux tends to \emph{smaller} present-day central temperatures, and hence, to relatively low initial temperatures (cold start).
At layer boundaries induced by a compositional gradient on the other hand, the temperature rises faster than in the adiabatic case leading to \emph{higher} present-day central temperatures. 
In that case, heat from the initial hot start remains restored in deep shells and is prevented from escaping to the surface reducing the total $L_{int}$. In this picture, the smaller intrinsic luminosity of Uranus arises from a more extended convectively stable region or from a colder start compared with Neptune. Both possibilities can potentially be explained by different characteristics of giant impacts during formation.  (See~\citet{Nbook} for a detailed discussion.) Furthermore, both possibilities are not in contradiction to the apparent similarity of the interior models presented in \S~\ref{ssec:UN_McZZ}, since $J_2$ and $J_4$ are not unique with respect to the density distribution on small scales.

In \ref{JF:sec:evolJS} we have seen that gravitational settling of immiscible material tends to lengthen the cooling time of Saturn by some gigayears; equivalently, redistribution of water from the inner envelope to the outer H/He envelope due to immiscibility offers an explanation for Uranus's low $T_{\rm eff}$. One step forward could be a calculation of cooling curves using non-adiabatic temperature gradients and heat transport through diffusive layers, and the calculation of material properties of  gas-ice-rock mixtures.

\section{Discussion}\label{sec:diss} 

\subsection{The concept of the core mass}\label{ssec:diss_core}

In sections \S~\ref{ssec:Jup_McMZ} we presented results for the core mass and metallicity of Jupiter (Uranus and Neptune: \S~\ref{ssec:UN_McZZ}) assuming a core composed of rocks or ices (U and N: 100\% rocks) and metals in the H/He envelopes being ice or ice-rock mixtures (U and N: H$_2$O). These approximations for Uranus and Neptune have been applied also by~\citet{FH03} on Saturn evolution models. Other Jupiter and Saturn models not presented here, e.g. by~\cite{Chabrier92}, assumed for the core a central agglomeration of rocks overlayed by an ice shell. Such assumptions can be considered \textit{state-of-the-art}.

\begin{figure}[hbt]
\includegraphics[angle=0,width=\textwidth]{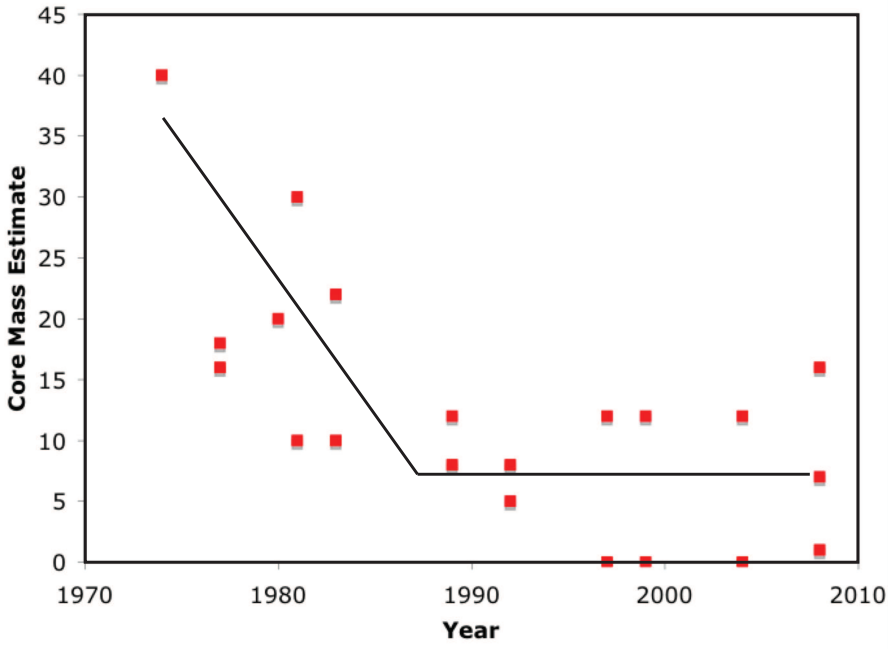}
\caption{Jupiter's core mass, as derived by many different authors, at various times since the early 1970s.}
\label{core}
\end{figure}

In Fig.~\ref{core} we show a collection of model derivations of Jupiter's core mass derived by a variety of authors over the past 35 years.  The spread is large.  Generally, as our understanding of H/He under high pressure has (presumably) improved, core masses have fallen.  Notably, in the 1970s and 1980s, a variety of groups used a variety of different H/He EOSs to compute structure models.  From the mid 1990s to mid 2000s, essentially only the \citet{SC95} EOS was used, predominantly by T.~Guillot.  We have now finally entered the era of first-principles calculations of H and He EOSs, and the behavior of this diagram over the coming years will be quite interesting.  Since the very nature of a well-behaved layered planet is only an assumption, in the following we also look at more complex diluted cores.  With gravity field data alone, it is not possible to differentiate between these simple and more complex models.

A common feature of Uranus and Neptune models is a large inner envelope metallicity, in our case up to 0.95 in mass, bringing it close to an ice shell. The small rocky core of Uranus and Neptune models, together with this almost-ice shell, resembles a large core. With $0-2M_{\oplus}$ central rocks and $9-12 M_{\oplus}$ of envelope H$_2$O in Uranus ($12-14.5 M_{\oplus}$ in Neptune), this gives a central mass of heavy elements of $\sim 11.5 M_{\oplus}$ for Uranus and $\sim 14.5 M_{\oplus}$ for Neptune, since larger rocky cores are accompanied by smaller $Z_2$ values. For brevity, we call this mass $M_{23,\,Z}$, the mass of the Z-component in layers 2 and 3. It is in good agreement with the core mass predicted by the core accretion formation models (CAF) models by~\citet{Pollack96}. More recent CAF models however by~\cite{Alibert05b} predict significantly smaller core masses of $\sim 6 M_{\oplus}$ for Jupiter and Saturn. Uranus' and Neptune's $M_{23,Z}$ is larger than Jupiter's $M_{core}$ (except if using DFT-MD, which gives $14-18 M_{\oplus}$). 
An obvious consequence is the following hypothesis:  All solar system giant planets formed by CAF with an initial core mass of $\sim 5-15 M_{\oplus}$. A deviation of their present core mass from this value indicates dissolving of initial core material within the deep interior, and does not indicate an inconsistency with CAF.

This dissolving of core material may have happened in the early hot stages of the planet's evolution or within a continuous, slowly progressing process. To explain Jupiter's relatively small derived core, \citet{SG04} suggest a larger mixing of core material in Jupiter than in Saturn due to a larger gas accretion rate during formation; in this sense, the high metallicity of Uranus' and Neptune's inner envelope implies weak core erosion and thus a small gas accretion rate in agreement with their small derived total gas fraction. 
A small Jupiter core today can also be explained by continuous, slow erosion. If the proto-core contained ice, this ice at present Jupiter core conditions of $\sim 20000$~K and $>40$~Mbar would be in the plasma phase~\citep{French09} which is soluble with hydrogen. However, we do not know how fast such an ice-enriched H/He/ice mixture can be redistributed by convection. Instead, a deep layer of H/He/ice can form which is stable against convection due to a compositional gradient. Note that an extended compositional gradient is not a preferred solution because of Jupiter's large heat flux, which strongly points to large-scale convection. 

\begin{figure}[hbt]
\includegraphics[angle=0,width=\textwidth]{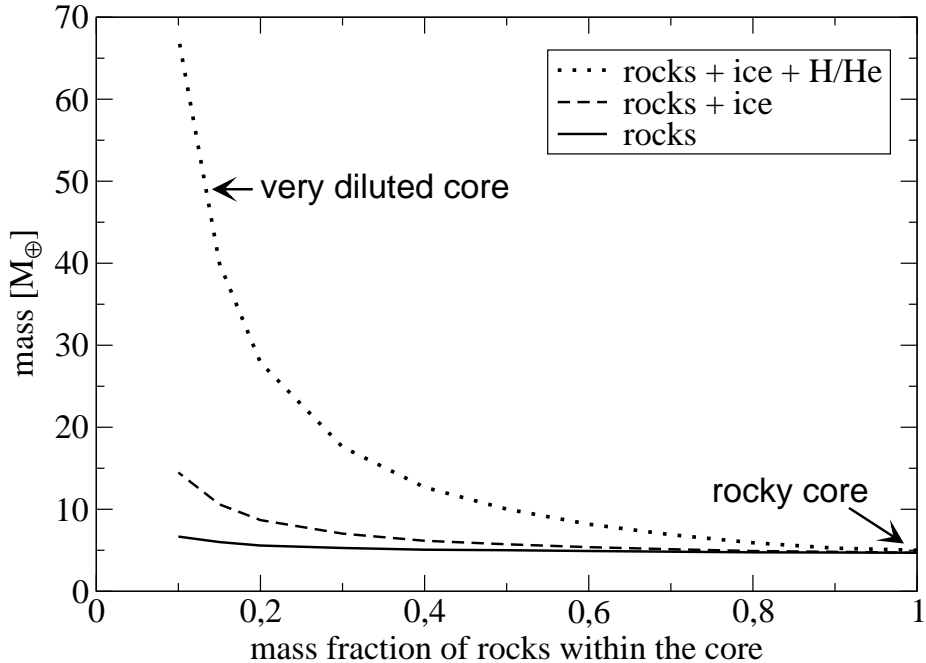}
\caption{Core mass of Jupiter assuming an isothermal core of H/He, water, and rocks with varying rock mass fraction. For all underlying models, the water to H/He mass ratio in the core is the same as in the inner envelope. The rock mass fraction in the core is varied between 100\% (usual rocky core) and 10\% (very diluted core). \textit{Solid line}: mass of rocks in the core, \textit{dashed line}: mass of rocks and water in the core, \textit{dotted line}: total core mass.}
\label{fig:dilutedCore}
\end{figure}

Within a simplified Jupiter model we can examine if a central region containing rocks, ice, and H/He can have $\sim 10 M_{\oplus}$ of heavy elements. For this examination we use LM-REOS. We assume a central region containing H/He and H$_2$O in the same relative fraction as in the usual deep envelope and vary the fraction of rocks in the central region. The result is shown in Fig.~\ref{fig:dilutedCore}.

It turns out that for rock mass fractions $X_{R,\,core}$ between 100 and 60\% in the central region,  the mass $M_{Z,\,core}$ of heavy elements is essentially unaffected. In order to obtain $M_{Z,\,core}>10 M_{\oplus}$ decreasing $X_{\rm R,\,core}$ down to $<20$\% is required. These models have $>30 M_{\oplus}$ H/He in the central region, the pressure at the core-mantle boundary decreases from 39 to 23~Mbar, and the core region growth from $\sim 1 R_{\oplus}$ to $>3 R_{\oplus}$.
The larger core region tends to enhance $J_2$ which in turn forces the fitting procedure to smaller inner envelope metallicities $Z_2$. In order to keep $|J_4|$ at a constant value, which decreases with  smaller $Z_2$, $Z_1$ must become larger by some $\Delta Z_1$. For $X_{R, \,core}<0.2$ we find $\Delta Z_1>50\%$. This should be kept in mind when evaluating models obtained with different EOS as presented in \S~\ref{ssec:Jup_McMZ}.

\subsection{Summary and Conclusions}

\paragraph{}
Since the pioneering work of \citet{DeMarcus58} over 50 years ago, it has been clear that Jupiter is composed predominantly of H and He.  But its content and distribution of heavy elements is still a matter of debate, despite great efforts to precisely measure its gravity field and huge advances in high-pressure experiments for H.

On the observational side, the unknown extent of differential rotation into the interior has given room to a variety of re-interpretations of the measured $J_4$ value. $J_4$ is an important quantity that strongly influences the distribution of metals in the envelope.  Whether or not homogeneous envelope models are consistent with $J_4$ depends on the EOS.  Accurate higher order moments from the \emph{Juno} mission might greatly advance our understanding of Jupiter's differential rotation, thereby constraining interior models.

Neglecting differential rotation, $J_2$ and $J_4$ and the EOS allow one to restrict Jupiter's core mass to $0-7 M_{\oplus}$ and the envelope metallicity to $11-14 M_{\oplus}$; including  differential rotation, this uncertainty rises to $M_{core}=0-18 M_{\oplus}$ and $M_Z=2-37 M_{\oplus}$ with $M_{core}+M_Z=12-37 M_{\oplus}$.  

\paragraph{{\rm For} Uranus and Neptune} we obtain a deep envelope metallicity of 80-95\%. Larger fractions of rock (or ices lighter than H$_2$O) would shift this range towards smaller (higher) values. These models resemble a slightly eroded ice-rock core of $\sim11 M_{\oplus}$~(U) and $\sim 15 M_{\oplus}$~(N) below a thin, ice-enriched H/He layer.

\paragraph{Eroded core models} of Jupiter give ice-rock core masses below $10 M_{\oplus}$ unless the core is assumed to be very diluted. This would indicate partial redistribution of core material into Jupiter's envelope. Any prediction of Jupiter's formation process from its present core mass is highly unreliable.

\section{Exoplanets} \label{exo}

\subsection{Current Explanations for Large Radii of Gas Giants} \label{rad}
\begin{figure}
  \includegraphics[width=1.0\textwidth]{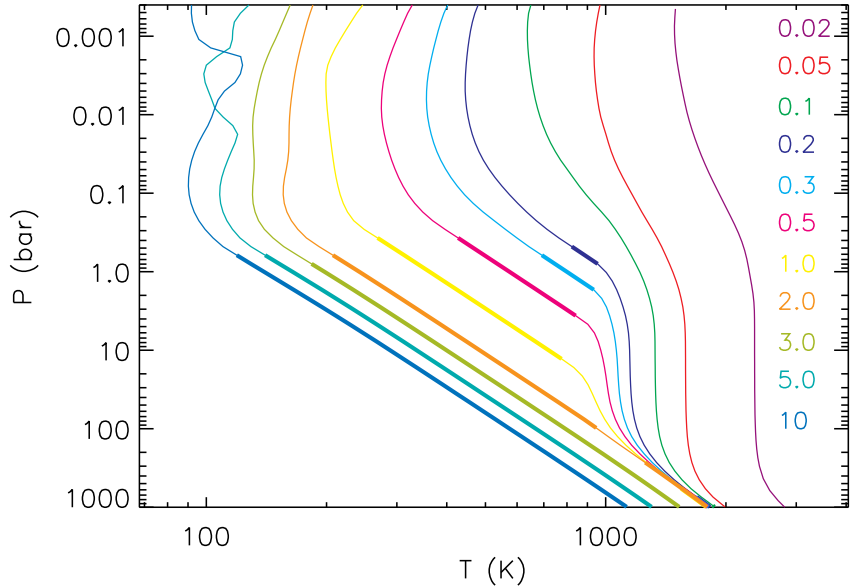}
\caption{(Color online)  Pressure-temperature profiles for 4.5 Gyr Jupiter-like planets ($g$=25 ms$^{-2}$, $T_{\rm int}=$ 100 K) from 0.02 to 10 AU from the Sun. Distance from the Sun in AU is color coded along the right side of the plot. Thick lines are convective regions, while thin lines are radiative regions.  Planets closer to the Sun have deeper atmospheric radiative zones.  The profiles at 5 and 10 AU show deviations that arise from numerical noise in the chemical equilibrium table near condensation points, but this has a negligible effect on planetary evolution.}
\label{pt}       
\end{figure}

As discussed in early sections, the standard cooling theory for giant planets \cp[e.g.,][]{Hubbard02} envisions an adiabatic H/He envelope, likely enhanced in heavy elements, on top of a distinct heavy element core, likely composed of ices and rocks.  It is the radiative atmosphere that serves as the bottleneck for interior cooling and contraction.  The effects of modest Jovian-like stellar irradiation on cooling models of Jupiter was investigated by \ct{Hubbard77}.  The ways in which strong stellar irradiation retards the contraction and interior cooling of giant planets was first worked out by \ct{Guillot96}.  The high external radiation keeps the atmosphere quite hot (1000-2000 K) and drives a shallow radiative temperature gradient deep into the atmosphere, to pressures of $\sim$1 kbar.  A shallower $dT/dP$ gradient in the atmosphere, compared to an isolated planet, means that the flux carried through the atmosphere must be necessarily reduced.  Atmospheric pressure-temperature profiles at a variety of incident flux levels are shown in Figure \ref{pt}.  Note that this incident flux itself does not directly effect the interior of the planet---the stellar flux is calculated to be wholly absorbed at pressures less than $\sim$5 bar \cp{Iro05}.  This means that these planet must reside in close-in orbits for their entire lives.  If they had previously cooled at 5 AU, and were brought in very recently, their radii would be $\sim$1 \rj, similar to Jupiter, with a very small increase in radius just due to a puffed up atmosphere \cp{Burrows00}.

\begin{figure}
  \includegraphics[width=1.0\textwidth]{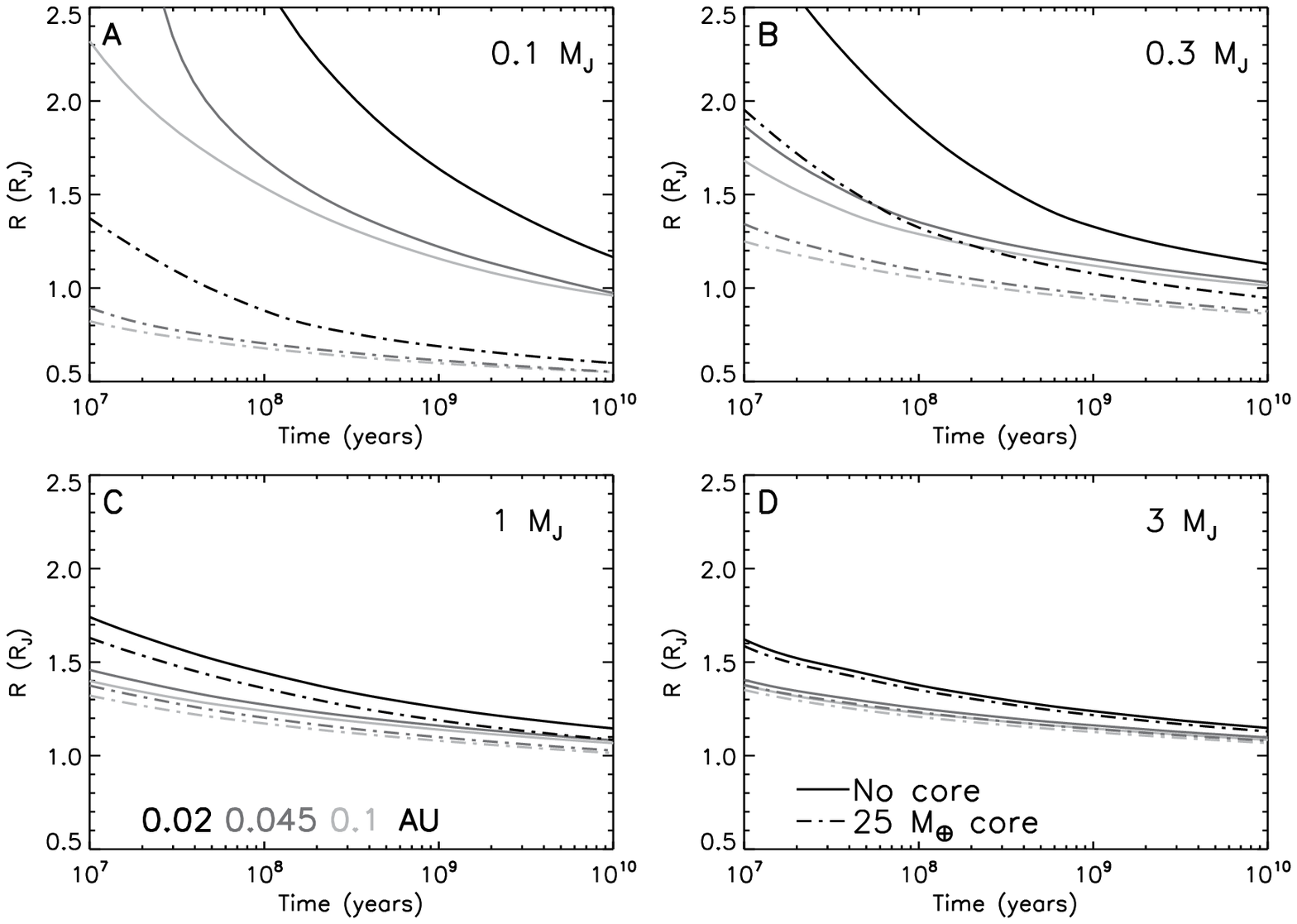}
\caption{Planetary radii as a function of time for masses of 0.1 \mj~(32 \me, $A$), 0.3 \mj~($B$), 1.0 \mj~($C$), and 3.0 \mj~($D$).  The three shades code for the three different orbital separations from the Sun, shown in ($C$).  Solid lines indicate models without cores and dash-dot lines indicate models with a core of 25 \me.}
\label{quad}       
\end{figure}

\begin{figure*}
  \includegraphics[width=1.0\textwidth]{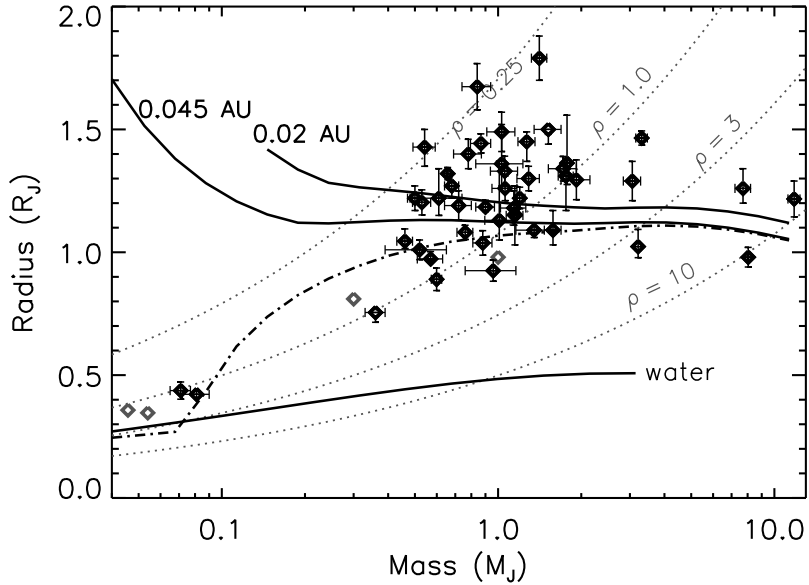}
\caption{The masses and radii of known transiting exoplanets, with error bars.  Planetary radii from the models of \ct{Fortney07a} at 4.5 Gyr.  The solid-curve models are for planets at 0.02 AU and 0.045 AU, with a composition of pure H/He (which is likely unrealistic given the structure of Jupiter \& Saturn), and includes the transit radius effect.  Models with a dash-dot curve at 0.045 AU include 25 \me\ of heavy elements (50/50 ice and rock) in a distinct core.  The lower solid curve is for pure water planets.  Diamonds without error bars are solar system planets.}
\label{mr}       
\end{figure*}

The upshot of this shallow atmospheric temperature gradient is that a smaller flux from the deep interior can be carried through the atmosphere---the cooling of the interior (and hence, contraction) is slowed, compared to the isolated case \cp{Guillot02,Baraffe03,Burrows03}.  The effects of irradiation of 0.02, 0.045, and 0.1 AU from a constant luminosity Sun are shown in Figure \ref{quad}, using the models of \ct{Fortney07a}.  For Jupiter-mass planets, radii of 1.2 \rj\ are expected at gigayear ages.  Nevertheless, as seen in \mbox{Figure~\ref{mr}}, many planets have radii in excess of 1.2 \rj, and most receive irradiation far below that expected at 0.02 AU.  Explaining the large radii has been a major focus of exoplanet research for several years.  Below we briefly review the previous work.
\begin{itemize}
\item{\ct{Bodenheimer01} proposed that the radius of \hd\ could be explained by non-zero orbital eccentricity, forced by an unseen additional planetary companion.  This eccentricity would then be tidally damped, perhaps for gigayears.  For \hd\ and other planets, this is potentially ruled out by the timing of the secondary eclipse \cp[e.g.,][]{Deming05b}, which indicates an eccentricity of zero.  Interest in tides continues, however.  \ct{Jackson08a} have shown the orbits of hot Jupiters are still decaying due to the tide raised on the star by the planet, and that tidal heating in the not-to-distant past could have been appreciable \cp{Jackson08}.  \ct{Levrard09} have followed up on this work and shown that nearly all detected transiting planets will eventually fall into their parent stars.  Recently \ct{Ibgui09} and \ct{Miller09} have extended the \ct{Jackson08} work by computing the first hot Jupiter contraction models that explicitly couple tidal heating to the thermal evolution of giant planets.  While tidal heating should be very important for some systems, it likely cannot explain all of the inflated planets.}
\item{\ct{Guillot02} proposed that a small fraction ($\sim$0.5-1\%) of absorbed stellar flux is converted to kinetic energy (winds) and dissipated at a depth of tens of bars by, e.~g., the breaking of atmospheric waves.  This mechanism would presumably effect all hot Jupiters to some degree.  While this mechanism is attractive, much additional work is needed to develop it in detail, as \ct{Burkert05} did not find this dissipation in their simulations.  Direct simulation of these atmospheres in 3D over long time scales is computationally expensive.}
\item{\ct{Baraffe04} found that \hd\ could be in the midst of extreme envelope evaporation, leading to a large radius, and we are catching the planet at a special time in its evolution.  The authors themselves judged this to be very unlikely.  Current models of atmospheric escape from hot Jupiters \cp[e.g.][]{Murray09} find evaporation rates much lower than those previously assumed by Baraffe and collaborators, which were based on earlier work.}
\item{\ct{Winn05} found that \hd\ may be stuck in a Cassini state, with its obliquity turned over at 90 degrees, which leads to a tidal damping of obliquity over gigayear ages.  Additional work by \ct{Levrard07} and \ct{Fabrycky07} have cast serious doubt on this mechanism for \hd\ and all close-in planets.  This work was recently reviewed in some detail by \ct{Peale08}.}
\item{\ct{Burrows07} propose that atmospheres with significantly enhanced opacities (10$\times$ that of a solar mixture) would stall the cooling and contraction of the planetary interior, leading to larger radii at gigayear ages.  This would be due to, for example, a large underestimation of the true opacities in these atmospheres \cp[see also][]{Ikoma06}.  Spectra of hot Jupiter atmospheres will either support or refute this (currently ad-hoc) possibility.  We note that if the H/He envelope were wholly 10$\times$ solar in \emph{metallicity}, the increased molecular weight of the H/He-dominated envelope would entirely negate this high-opacity effect \cp{Hansen07,Guillot08}.}
\item{\ct{Chabrier07c}, independently following along the lines of a hypothesis from \ct{Stevenson85}, suggest that gradients in heavy elements (such as from core dredge-up or dissolution of planetesimals) could suppress convection and cooling in the H/He envelope, leading to large radii at gigayear ages.  This double diffusive convection (where there are gradients in both temperature and composition) occurs in the Earth's oceans.  These diffusive layers could, however, be quite fragile, and 3D simulations of this process are required, under conditions relevant to giant planet interiors.  Note that this effect could be present for planets at any orbital distance.}
\item{\ct{Hansen07} suggest that if mass loss due to evaporation leads to a preferential loss of He vs.~H (perhaps due to magnetic fields confining H$^+$), that planets could be larger than expected due to a smaller mean molecular weight.  This mechanism would also presumably effect all hot Jupiters to some degree.  However, \ct{Guillot08} has shown that some planets are still larger than can be accommodated by pure hydrogen composition.}
\item{\ct{Arras09} very recently postulated that a thermal tide in the atmosphere of hot Jupiters could lead to energy dissipation in their atmospheres, thereby potentially leading to inflated radii.  However, there appear to be problems with the implementation in this work, and \ct{Gu09} perform a somewhat similar analysis and find very weak energy dissipation.}

\end{itemize}

We note that a planetary radius-inflation mechanism that would affect all hot Jupiters is quite reasonable.  Since giant planets are expected to be metal rich (Jupiter and Saturn are 5-20\% heavy elements) a mechanism that would otherwise lead to large radii could easily be canceled out by a large planetary core or a supersolar abundance of heavy elements in the H/He envelope in most planets \cp{Fortney06}.  Planets that appear ``too small'' are certainly expected and are relatively easy to account for due to a diversity in internal heavy element abundances \cp{Guillot06,Burrows07}.

Some of these inflation mechanisms should scale with stellar irradiation level or with orbital separation, while others do not.  Therefore, a premium should be placed on finding transiting planets farther from their parent stars.  All but two of the known transiting planets have orbits of only 1-6 days.  However, the French \emph{CoRoT} and American \emph{Kepler} missions have the potential to find transiting giant planets out to 0.2 AU and 1 AU, respectively.  \emph{CoRoT} has already announced 5 planets in close-in orbits, and \emph{Kepler} just launched in March 2009.  The orbital separation limits on these missions are due entirely to the length of time these telescopes will stare at a given patch of sky---the longer the time duration, the longer the planetary orbital period that can be seen to have multiple transits.  This science will continue to expand, and the future is bright.

\subsection{The Expanding Field of Exo-Neptunes}
The transits of planets GJ 436b \cp{Gillon07} and HAT-P-11b \cp{Bakos09} have opened the field of direct characterization of Neptune-class planets in addition to Jupiter- and Saturn-class.  This is extremely exciting.  Two things that we have immediately learned from merely a measured mass and radius are that: 1) These planets \emph{must} have H/He envelopes (they cannot be purely heavy elements), but that these envelopes are probably only 10-20 \% of the planet's mass, similar to Uranus and Neptune.  2) That these two planets are not likely to be remnants of evaporated gas giants.  \ct{Baraffe06} had calculated that Neptune-mass planets that are evaporation remnants should have large radii around $\sim$1 \rj, due to a tenuous remaining gaseous envelope, while these two planets have radii less than 0.5 \rj.  \ct{Hubbard07b} have also shown that the mass function of observed radial velocity exoplanets is nearly independent of orbital distance.  If evaporation were important, one might expect a deficit of close-in Saturn-mass planets, which would be easier to evaporate than more massive giants.  However, it will take a statistically interesting number of transiting planet detections before we can claim to see trends in these lower mass planets.

\subsection{Young Gas Giant Planets} \label{youngj} 
As discussed in the previous section, there is in wide use a model for the cooling and contraction of gas giant planets that is now being tested in a variety of cases at Gyr ages.  It is clear from giant planet formation theories that these planets are hot, luminous, and have larger radii at young ages, and they contract and cool inexorably as they age.  However, since the planet formation process is not well understood \emph{in detail}, we understand very little about the initial conditions for the planets' subsequent cooling.  Since the Kelvin-Helmholtz time is very short at young ages (when the luminosity is high and radius is large) it is expected that giant planets forget their initial conditions quickly.  This idea was established with the initial Jupiter cooling models in the 1970s \cp{Graboske75,Bodenheimer76}.

Since our solar system's giant planets are thought be 4.5 Gyr old, there is little worry about how thermal evolution models of these planets are effected by the unknown initial conditions.  The same may not be true for very young planets, however.  Since giant planets are considerably brighter at young ages, searches to directly image planets now focus on young stars.  At long last, these searches are now bearing fruit \cp{Chauvin05,Marois08,Kalas08}.  It is at ages of a few million years where understanding the initial conditions and early evolution history is particularly important, if we are to understand these planets.  Traditional evolution models, which are applied to both giant planets and brown dwarfs, employ an arbitrary starting point.  The initial model is large in radius, luminosity, and usually fully adiabatic.  The exact choice of the starting model is usually thought to be unimportant, if one is interested in following the evolution for ages greater than 1 Myr \cp{Burrows97,Chabrier00}.

We will now briefly discuss how these models are used.  Thermal evolution models, when coupled to a grid of model atmospheres, aim to predict the luminosity, radius, \teff, thermal emission spectrum, and reflected spectrum, as a function of time.  When a planetary candidate is imaged, often only the apparent magnitude in a few infrared bands are known, at least initially.  If the age of the parent star can be estimated (itself a tricky task) then the observed infrared magnitudes can be compared with calculations of model planets for various masses, to estimate the planet's mass, which is not an observable quantity unless some dynamical information is also known.  It is not known if these thermal evolution models are accurate at young ages--they are relatively untested, which has been stressed by \ct{Baraffe02} for brown dwarfs and \ct{Marley07} for planets.  Indeed, \ct{Stevenson82b} had stressed that these cooling models ``\dots cannot be expected to provide accurate information on the first $10^5-10^8$ years of evolution because of the artificiality of an initially adiabatic, homologously contracting state"

\begin{figure}
 \includegraphics[width=1.0\textwidth]{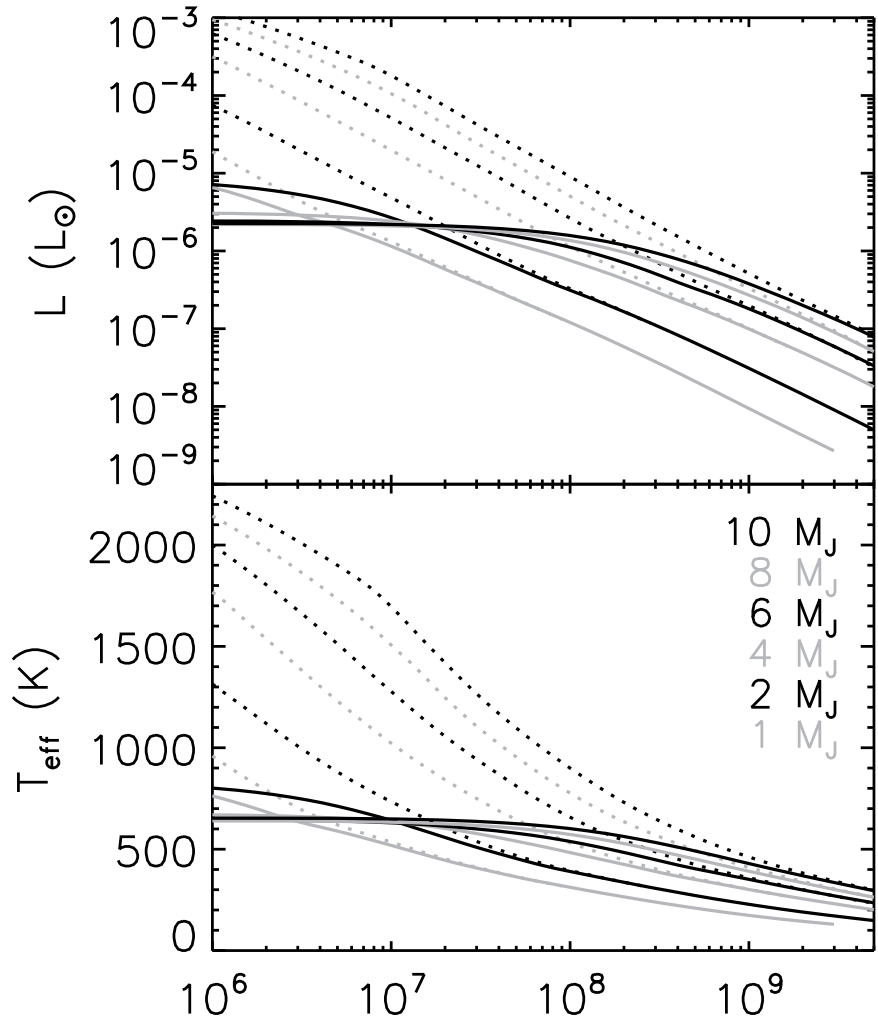}
  \caption{Models from \ct{Marley07} of the thermal evolution of giant planets from 1 to 10 \mj.  The dotted curves are standard ``hot start'' models with an arbitrary initial condition, and the solid curves use as an initial condition the core accretion formation models of \ct{Hubickyj05}.
}  \label{young}
\end{figure}

\ct{Marley07} examined the issue of the accuracy of the arbitrary initial conditions (termed a ``hot start'' by the authors) by using initial conditions for cooling that were not arbitrary, but rather were given by a leading core accretion planet formation model \cp{Hubickyj05}.  The core accretion calculation predicts the planetary structure at the end of formation, when the planet has reached its final mass.  The \ct{Marley07} cooling models use this initial model for time zero, and subsequent cooling was followed as in previously published models.  Figure \ref{young} shows the resulting evolution.  The cooling curves are dramatically different, yielding cooler (and smaller) planets.  The initial conditions are not quickly ``forgotten,'' meaning that the cooling curves do not overlap with the arbitrary start models for 10$^7$ to 10$^9$ years.  What this would mean, in principle, is that a mass derived from ``hot start'' evolutionary tracks would significantly underestimate the true mass of a planet formed by core accretion.

Certainly one must remember that a host of assumptions go into the formation model, which yields the starting point for evolution, so it is unlikely that these new models are quantitatively correct.  However, they highlight that much additional work is needed to understand the energetics of the planet formation process.  The \ct{Hubickyj05} models yield relatively cold initial planets because of an assumption that accreting gas is shocked and readily radiates away this energy.  The end result is that the accreted gas is of relatively low specific entropy, leading to a low luminosity starting point for subsequent evolution.  Significant additional work on multi-dimensional accretion must be done, as well as on radiative transfer during the accretion phase, before we can confidently model the early evolution.

Another issue, which is more model independent, is that since the planet formation by core accretion may take $\sim$1-5 Myr to complete, it is likely incorrect to assume that a parent star and its planets are coeval.  This will be particularly important for young systems.  If a planetary candidate with given magnitudes is detected, overestimating its age (since it would be younger than its parent star) would lead to an overestimation of its mass.  Thankfully, it appears that detections of young planets are now beginning to progress quickly, which will help to constrain these models.

\subsection{Conclusions: Exoplanets}
Since the information that we can gather about interiors of the solar system's giant planets is inherently limited, advances in understanding giant planets as classes of astronomical objects will likely rest on the characterization of a large number of exoplanets.  While for any particular planet, the amount of knowledge to be gleaned is relatively small, this can be overcome by the shear numbers of these planets.  Therefore, in the future, some of this work will necessarily have to be statistical in nature.  This has already begun to some degree, as \ct{Fressin07,Fressin09} have analyzed current transit surveys to derive the yields and giant planet properties of from these observations.

Understanding the mass-radius relation of giant planets as a function of orbital distance is a critically important question.  What is causing the large planetary radii and how does it scale with distance?  The French \emph{CoRoT} mission should be able to detect planets out to 0.2 AU, and the American \emph{Kepler} mission out to 1 AU, due to its longer time baseline.  Any planets found in these wider orbits will be critical data points.  After \emph{Kepler}, it is not at all clear when, \emph{if ever}, we may have access to precise radii and masses of giant planets for planets in orbits of months to years.

The direct imaging of giant planets is now ramping up and allows us to sample additional parameter space---mostly young, massive planets far from their parent stars.  Determining the physical properties of these planets in eras not long after their formation will allow us to better understand planet formation and thermal evolution.




\end{document}